\documentstyle[11pt,psfig]{article}
\begin{document}
\newcommand{\lton}{\stackrel{\large <}{\sim}}
\newcommand{\gton}{\stackrel{\large >}{\sim}}
\newcommand{\beq}{\begin{equation}}
\newcommand{\beqar}{\begin{eqnarray}}
\newcommand{\eeq}[1]{\label{#1} \end{equation}}
\newcommand{\eeqar}[1]{\label{#1} \end{eqnarray}}
\newcommand{\gfm}{{\rm GeV/Fm}^3}
\newcommand{\fsbmn}{\langle F_{\mu \nu} \rangle}
\newcommand{\fspmn}{\langle F^{\mu \nu} \rangle}
\newcommand{\asbm}{\langle A_\mu \rangle}
\newcommand{\aspm}{\langle A^\mu \rangle}
\newcommand{\epsi}{\vec{\epsilon}}
\textfloatsep=0.2in
\renewcommand{\topfraction}{0.99}
\renewcommand{\bottomfraction}{0.99}
\renewcommand{\textfraction}{0.01}
\setcounter{topnumber}{1}
\pagestyle{myheadings}
\thispagestyle{empty}
\pagestyle{myheadings}
\markboth
{\it B. Pfeiffer, K.-L. Kratz, P. M\"{o}ller,  Delayed Neutron-Emission Probabilities \dots}
{\it B. Pfeiffer, K.-L. Kratz, P. M\"{o}ller,  Delayed Neutron-Emission Probabilities \dots}
\mbox{ } \\
\mbox{ } \vspace{-0.90in}\mbox{ }\\
\begin{center}
\begin{Large}
\begin{bf}
STATUS OF DELAYED-NEUTRON PRECURSOR DATA:\\
HALF-LIVES AND NEUTRON EMISSION PROBABILITIES\\[2ex]

\end{bf}
\end{Large}
Bernd Pfeiffer\footnote{E-mail address: Bernd.Pfeiffer@uni-mainz.de} 
 and Karl-Ludwig Kratz\\
{\it Institut f\"ur Kernchemie, Universit\"at Mainz, Germany}\\[1ex]
Peter M\"{o}ller\\ 
{\it Theoretical Division, Los Alamos National Laboratory, Los Alamos, NM 87545}\\[1ex]
\end{center}

\begin{description}
\sloppy
\item[{\large\sc Abstract: --}]
We present in this paper a compilation of the present status 
of experimental delayed-neutron precursor data; i.e. $\beta$-decay half-lives 
($T_{1/2}$) and neutron emission probabilities ($P_{\rm n}$) in the 
fission-product region ($27 \le Z \le 57$). 
These data are compared to two model predictions of substantially different 
sophistication: (i) an update of the empirical Kratz--Herrmann formula (KHF), 
and (ii) a unified macroscopic-microscopic model within the quasi-particle 
random-phase approximation (QRPA). Both models are also used to calculate so 
far unknown $T_{1/2}$ and $P_{\rm n}$ values up to $Z=63$. A number of 
possible refinements in the microscopic calculations are suggested to further 
improve the nuclear-physics foundation of these data for reactor
and astrophysical applications.
\end{description}

\begin{center}
INTRODUCTION
\end{center}

Half-lives ($T_{1/2}$) and delayed-neutron emission probabilities 
($P_{\rm n}$) are 
among the easiest measurable gross $\beta$-decay properties of neutron-rich 
nuclei far from stability. They are not only of importance for reactor 
\begin{figure}[t]
\vspace{1cm}
\centerline{\psfig{file=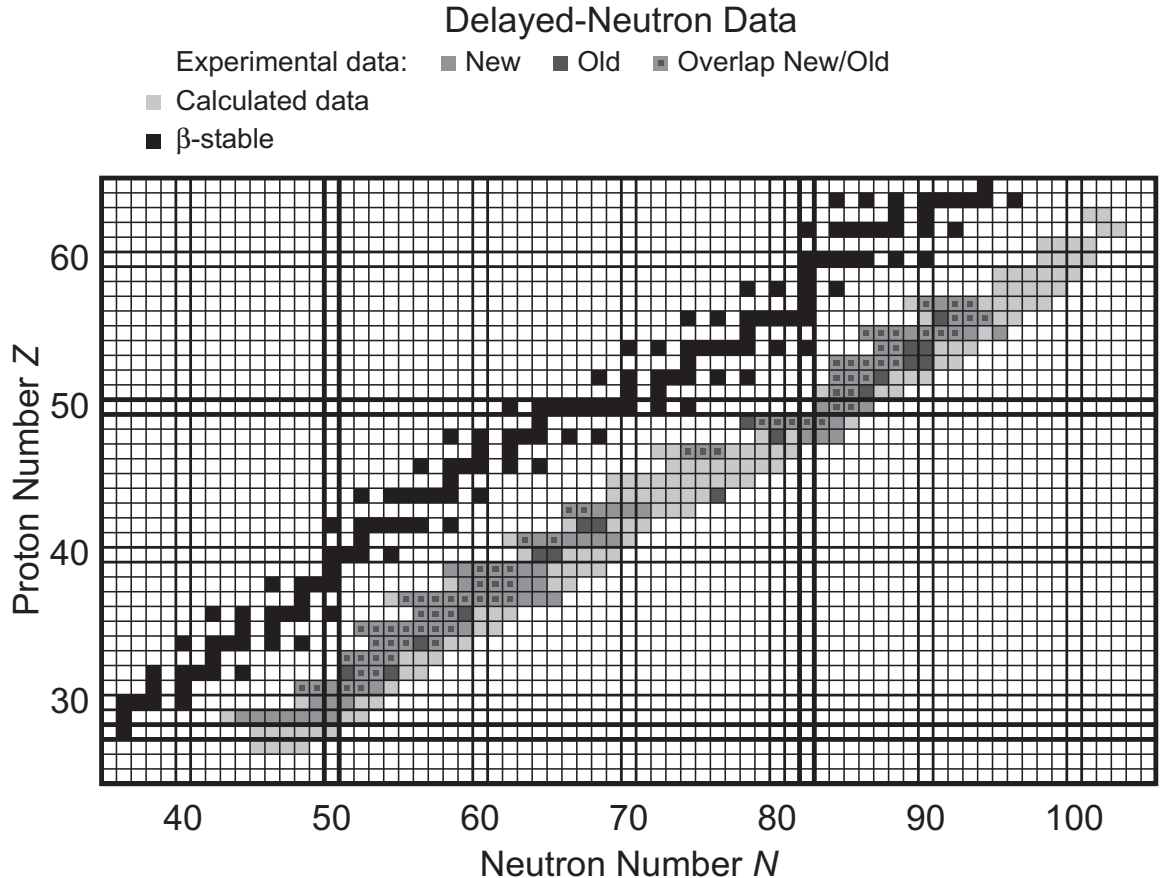,angle=-90,width=6.0in}}
\caption[chart]{Chart illustrating the data available in the fission-product
region. The new data evaluation represents a significant extension of
measured $P_{\rm n}$ values. Some data in the old data set
are not present in the new data set.}
\label{chart}
\end{figure}
applications, but also in the context of studying nuclear-structure features 
and astrophysical scenarios. Therefore, most of our recent experiments 
performed at international facilities such as CERN-ISOLDE, GANIL-LISE and 
GSI-FRS were primarily motivated by our current work on r-process 
nucleosynthesis. However, it is a pleasure for us to recognize that these 
data still today may be of interest for applications in reactor physics, a 
field which we practically left shortly after the ''Specialists' Meeting on 
Delayed Neutrons'' held at Birmingham in 1986. \\

Our motivation to put together this new compilation of
$\beta$-decay half-lives and $\beta$-delayed neutron-emission
 came from recent 
discussions with T.R. England and W.B. Wilson from LANL about our activities 
in compiling and steadily updating experimental delayed-neutron data as well 
as various theoretical model predictions (Pfeiffer {\it et al.}, 2000).
They pointed out to us, that their recent
summation calculations of aggregate fission-product delayed-neutron production 
using basic nuclear data from the early 1990's (Brady, 1989;
Brady and England, 1989, Rudstam 1993)
show, in general, that a greater fraction of delayed neutrons is emitted 
at earlier times following fission than measured. As a consequence, the 
reactor response to a reference reactivity change is enhanced compared to that 
calculated with pulse functions derived from measurements (Wilson and England, 
2000).
Therefore, the use of updated $P_{\rm n}$ and 
$T_{1/2}$ values is expected to improve the physics foundation of the basic 
input data used and to increase the accuracy of aggregate results obtained in 
summation calculations.\\

Since the tabulation of Brady (1989) and Rudstam (1993),
about 40 new $P_{\rm n}$ values have been measured in the fission-product region
($27\le Z\le 57 $), a number of delayed-neutron branching ratios have also been 
determined with higher precision, and a similar number of ground-state
and isomer decay half-lives of new delayed-neutron precursors have been obtained. These 
data are contained in our compilation (Table~\ref{table1}), and are compared 
with two of 
our model predictions: (i) an update of the empirical Kratz-Herrmann 
formula (KHF) for $\beta$-delayed neutron emission probabilities $P_{\rm n}$ 
and $\beta$-decay half-lives $T_{1/2}$
(Kratz and Herrmann, 1973; Pfeiffer, 2000),
and an improved version of the 
macroscopic-microscopic QRPA model (M\"oller and Randrup, 1990)
which can be used to calculate a large number of nuclear properties 
consistently (M\"oller {\it et al.}, 1997).
These two models, with  quite different nuclear-structure basis,
 are also used to predict so far unknown $T_{1/2}$ and $P_{\rm n}$ 
values in the fission-product region (see Table~\ref{table1}). \\

\begin{table}
\caption[tabmas]{
Experimental $\beta$-decay half-lives $T_{1/2}$
and $\beta$-delayed neutron-emission probabilities $P_{\rm n}$
\label{table1}compared to three calculations.}
\begin{center}
\begin{tabular}{r|rrrrr|rrrrr}
\hline
& \multicolumn{5}{|c|}{$T_{1/2}$ (ms)}
&\multicolumn{5}{c}{$P_{\rm n}$  (\%)} \\
\hline
Isotope & \multicolumn{2}{|c}{Exp.} & KHF & QRPA-1 & QRPA-2
& \multicolumn{2}{|c}{Exp.} & KHF & QRPA-1 & QRPA-2\\
\hline
$^{ 66}_{27}$Co & 
       180 &         10 &       432 &       260 &       260 & 
                     &                    &     0.013 &      0.000 &      0.000\\
$^{ 67}_{27}$Co & 
       425 &         20 &       400 &        76 &       120 & 
                     &                    &     1.132 &      0.156 &      0.061\\
$^{ 68}_{27}$Co & 
       230 &         30 &       168 &        72 &        82 & 
                     &                    &     2.395 &      0.812 &      0.768\\
$^{ 68{\rm m} }_{\phantom{{\rm m} }27}$Co & 
      1600 &        300 & & & & 
                     &                    & & &  \\
$^{ 69}_{27}$Co & 
       216 &          9 &       238 &        64 &        99 & 
                     &                    &     6.585 &      1.250 &      0.905\\
$^{ 70}_{27}$Co & 
       120 &         30 &       101 &        43 &        60 & 
                     &                    &     7.469 &      3.213 &      2.154\\
\hline
\end{tabular}\\[3ex]
\end{center}
\end{table}
\setcounter{table}{0}
\begin{table}
\caption[tabmas]{
Continued}
\begin{center}
\begin{tabular}{r|rrrrr|rrrrr}
\hline
& \multicolumn{5}{|c|}{$T_{1/2}$ (ms)}
&\multicolumn{5}{c}{$P_{\rm n}$  (\%)} \\
\hline
Isotope & \multicolumn{2}{|c}{Exp.} & KHF & QRPA-1 & QRPA-2
& \multicolumn{2}{|c}{Exp.} & KHF & QRPA-1 & QRPA-2\\
\hline
$^{ 80}_{30}$Zn & 
       545 &         16 &       255 &      3025 &      2033 & 
                 1.0 &                   5&     0.668 &     10.889 &      9.980\\
$^{ 81}_{30}$Zn & 
       290 &         50 &        64 &       646 &      2401 & 
                 7.5 &                  30&     2.343 &     21.952 &     60.505\\
$^{ 82}_{30}$Zn & 
           &            &        52 &       211 &       734 & 
                     &                    &    17.170 &     35.281 &     99.972\\
$^{ 83}_{30}$Zn & 
           &            &        43 &        22 &       818 & 
                     &                    &    23.300 &      9.816 &     99.982\\
$^{ 84}_{30}$Zn & 
           &            &        43 &        65 &       387 & 
                     &                    &    25.480 &     33.273 &    100.000\\
 & & & & &   & & & & & \\
\hline
\end{tabular}\\[3ex]
\end{center}
\end{table}
\setcounter{table}{0}
\begin{table}
\caption[tabmas]{
Continued}
\begin{center}
\begin{tabular}{r|rrrrr|rrrrr}
\hline
& \multicolumn{5}{|c|}{$T_{1/2}$ (ms)}
&\multicolumn{5}{c}{$P_{\rm n}$  (\%)} \\
\hline
Isotope & \multicolumn{2}{|c}{Exp.} & KHF & QRPA-1 & QRPA-2
& \multicolumn{2}{|c}{Exp.} & KHF & QRPA-1 & QRPA-2\\
\hline
$^{ 92}_{34}$Se & 
           &            &        93 &        59 &       164 & 
                     &                    &    11.770 &      2.187 &      1.312\\
$^{ 93}_{34}$Se & 
           &            &        62 &        87 &        24 & 
                     &                    &    10.910 &     24.270 &      9.342\\
$^{ 94}_{34}$Se & 
           &            &        59 &        52 &        43 & 
                     &                    &    13.830 &     19.996 &      1.712\\
 & & & & &   & & & & & \\[-0.12in]
$^{ 87}_{35}$Br & 
     55600 &        150 &      1059 &     12315 &     37254 & 
                2.52 &                   7&     0.146 &      0.277 &      1.129\\
$^{ 88}_{35}$Br & 
     16360 &         70 &       685 &      1386 &    104991 & 
                6.55 &                  18&     0.466 &      0.194 &     28.599\\
\hline
\end{tabular}\\[3ex]
\end{center}
\end{table}
\setcounter{table}{0}
\begin{table}
\caption[tabmas]{
Continued}
\begin{center}
\begin{tabular}{r|rrrrr|rrrrr}
\hline
& \multicolumn{5}{|c|}{$T_{1/2}$ (ms)}
&\multicolumn{5}{c}{$P_{\rm n}$  (\%)} \\
\hline
Isotope & \multicolumn{2}{|c}{Exp.} & KHF & QRPA-1 & QRPA-2
& \multicolumn{2}{|c}{Exp.} & KHF & QRPA-1 & QRPA-2\\
\hline
$^{102}_{\phantom{1}37}$Rb & 
        37 &          5 &        36 &        13 &        13 & 
                  18 &                   8&    25.270 &     20.312 &     20.312\\
$^{103}_{\phantom{1}37}$Rb & 
           &            &        39 &        17 &        17 & 
                     &                    &    48.740 &     52.410 &     52.410\\
$^{104}_{\phantom{1}37}$Rb & 
           &            &        27 &        13 &        13 & 
                     &                    &    37.980 &     58.332 &     58.332\\
$^{105}_{\phantom{1}37}$Rb & 
           &            &        28 &        12 &        12 & 
                     &                    &    55.730 &     73.481 &     73.481\\
 & & & & &   & & & & & \\[-0.12in]
$^{ 96}_{38}$Sr & 
      1070 &         10 &       854 &      1079 &       817 & 
                     &                    &     0.000 &      0.000 &      0.000\\
\hline
\end{tabular}\\[3ex]
\end{center}
\end{table}
\setcounter{table}{0}
\begin{table}
\caption[tabmas]{
Continued}
\begin{center}
\begin{tabular}{r|rrrrr|rrrrr}
\hline
& \multicolumn{5}{|c|}{$T_{1/2}$ (ms)}
&\multicolumn{5}{c}{$P_{\rm n}$  (\%)} \\
\hline
Isotope & \multicolumn{2}{|c}{Exp.} & KHF & QRPA-1 & QRPA-2
& \multicolumn{2}{|c}{Exp.} & KHF & QRPA-1 & QRPA-2\\
\hline
$^{107}_{\phantom{1}40}$Zr & 
           &            &       144 &       197 &       197 & 
                     &                    &     1.727 &      1.457 &      1.457\\
$^{108}_{\phantom{1}40}$Zr & 
           &            &       130 &       181 &       181 & 
                     &                    &     5.820 &      1.796 &      1.796\\
$^{109}_{\phantom{1}40}$Zr & 
           &            &       117 &       122 &       122 & 
                     &                    &     3.968 &      4.007 &      4.007\\
$^{110}_{\phantom{1}40}$Zr & 
           &            &        98 &        86 &        86 & 
                     &                    &     7.114 &      5.979 &      5.979\\
 & & & & &   & & & & & \\[-0.12in]
$^{103}_{\phantom{1}41}$Nb & 
      1500 &        200 &      3192 &      9535 &      9535 & 
                     &                    &     0.000 &      0.000 &      0.000\\
\hline
\end{tabular}\\[3ex]
\end{center}
\end{table}
\setcounter{table}{0}
\begin{table}
\caption[tabmas]{
Continued}
\begin{center}
\begin{tabular}{r|rrrrr|rrrrr}
\hline
& \multicolumn{5}{|c|}{$T_{1/2}$ (ms)}
&\multicolumn{5}{c}{$P_{\rm n}$  (\%)} \\
\hline
Isotope & \multicolumn{2}{|c}{Exp.} & KHF & QRPA-1 & QRPA-2
& \multicolumn{2}{|c}{Exp.} & KHF & QRPA-1 & QRPA-2\\
\hline
$^{116}_{\phantom{1}44}$Ru & 
           &            &       556 &       612 &       612 & 
                     &                    &     0.053 &      0.002 &      0.002\\
$^{117}_{\phantom{1}44}$Ru & 
           &            &       237 &       175 &       175 & 
                     &                    &     0.287 &      0.369 &      0.369\\
$^{118}_{\phantom{1}44}$Ru & 
           &            &       287 &       233 &       233 & 
                     &                    &     1.029 &      1.120 &      1.120\\
$^{119}_{\phantom{1}44}$Ru & 
           &            &       162 &       185 &       185 & 
                     &                    &     1.712 &      2.616 &      2.616\\
$^{120}_{\phantom{1}44}$Ru & 
           &            &       149 &       118 &       118 & 
                     &                    &     2.599 &      2.945 &      2.945\\
 & & & & &   & & & & & \\
\end{tabular}\\[3ex]
\end{center}
\end{table}
\setcounter{table}{0}
\begin{table}
\caption[tabmas]{
Continued}
\begin{center}
\begin{tabular}{r|rrrrr|rrrrr}
\hline
& \multicolumn{5}{|c|}{$T_{1/2}$ (ms)}
&\multicolumn{5}{c}{$P_{\rm n}$  (\%)} \\
\hline
Isotope & \multicolumn{2}{|c}{Exp.} & KHF & QRPA-1 & QRPA-2
& \multicolumn{2}{|c}{Exp.} & KHF & QRPA-1 & QRPA-2\\
\hline
$^{129{\rm m} }_{\phantom{1{\rm m} }47}$Ag & 
           &            & & & & 
                     &                    & & &  \\
$^{130}_{\phantom{1}47}$Ag & 
           &          &        30 &        36 &        36 & 
                     &                    &    19.240 &     67.219 &     67.219\\
$^{131}_{\phantom{1}47}$Ag & 
           &            &        28 &        40 &        40 & 
                     &                    &    68.150 &    100.000 &    100.000\\
$^{132}_{\phantom{1}47}$Ag & 
           &            &        20 &        34 &        34 & 
                     &                    &    61.090 &    100.000 &    100.000\\
 & & & & &   & & & & & \\[-0.12in]
$^{126}_{\phantom{1}48}$Cd & 
       506 &         15 &       798 &      5146 &      5146 & 
                     &                    &     0.000 &      0.000 &      0.000\\
\hline
\end{tabular}\\[3ex]
\end{center}
\end{table}
\setcounter{table}{0}
\begin{table}
\caption[tabmas]{
Continued}
\begin{center}
\begin{tabular}{r|rrrrr|rrrrr}
\hline
& \multicolumn{5}{|c|}{$T_{1/2}$ (ms)}
&\multicolumn{5}{c}{$P_{\rm n}$  (\%)} \\
\hline
Isotope & \multicolumn{2}{|c}{Exp.} & KHF & QRPA-1 & QRPA-2
& \multicolumn{2}{|c}{Exp.} & KHF & QRPA-1 & QRPA-2\\
\hline
$^{136}_{\phantom{1}49}$In & 
           &            &        30 &        69 &        69 & 
                     &                    &    56.880 &    100.000 &    100.000\\
$^{137}_{\phantom{1}49}$In & 
           &            &        31 &        48 &        48 & 
                     &                    &    79.100 &    100.000 &    100.000\\
 & & & & &   & & & & & \\[-0.12in]
$^{133}_{\phantom{1}50}$Sn & 
      1450 &         30 &       362 &      9479 &      9479 & 
              0.0294 &                  24&     0.002 &      0.040 &      0.040\\
$^{134}_{\phantom{1}50}$Sn & 
      1120 &         80 &       245 &      2196 &      2196 & 
                  17 &                  14&     6.000 &     93.128 &     93.128\\
$^{135}_{\phantom{1}50}$Sn & 
       450 &         50 &       215 &      2789 &      2789 & 
                  22 &                   7&    12.650 &     98.591 &     98.591\\
\hline
\end{tabular}\\[3ex]
\end{center}
\end{table}
\setcounter{table}{0}
\begin{table}
\caption[tabmas]{
Continued}
\begin{center}
\begin{tabular}{r|rrrrr|rrrrr}
\hline
& \multicolumn{5}{|c|}{$T_{1/2}$ (ms)}
&\multicolumn{5}{c}{$P_{\rm n}$  (\%)} \\
\hline
Isotope & \multicolumn{2}{|c}{Exp.} & KHF & QRPA-1 & QRPA-2
& \multicolumn{2}{|c}{Exp.} & KHF & QRPA-1 & QRPA-2\\
\hline
$^{142}_{\phantom{1}53}$I  & 
           &            &       308 &       182 &      1400 & 
                     &                    &    10.750 &     46.997 &     99.952\\
$^{143}_{\phantom{1}53}$I  & 
           &            &       296 &       150 &       150 & 
                     &                    &    21.460 &     77.120 &     77.120\\
$^{144}_{\phantom{1}53}$I  & 
           &            &       194 &        58 &        58 & 
                     &                    &    17.370 &     29.379 &     29.379\\
$^{145}_{\phantom{1}53}$I  & 
           &            &       127 &        57 &        57 & 
                     &                    &    38.580 &     46.359 &     46.359\\
$^{146}_{\phantom{1}53}$I  & 
           &            &        80 &        29 &        29 & 
                     &                    &    24.200 &     27.584 &     27.584\\
$^{147}_{\phantom{1}53}$I  & 
           &            &        75 &        33 &        33 & 
                     &                    &    41.940 &     59.854 &     59.854\\
\hline
\end{tabular}\\[3ex]
\end{center}
\end{table}
\setcounter{table}{0}
\begin{table}
\caption[tabmas]{
Continued}
\begin{center}
\begin{tabular}{r|rrrrr|rrrrr}
\hline
& \multicolumn{5}{|c|}{$T_{1/2}$ (ms)}
&\multicolumn{5}{c}{$P_{\rm n}$  (\%)} \\
\hline
Isotope & \multicolumn{2}{|c}{Exp.} & KHF & QRPA-1 & QRPA-2
& \multicolumn{2}{|c}{Exp.} & KHF & QRPA-1 & QRPA-2\\
\hline
$^{149}_{\phantom{1}56}$Ba & 
       344 &          7 &       467 &       300 &       300 & 
                0.79 &                  39&     0.092 &      0.123 &      0.123\\
$^{150}_{\phantom{1}56}$Ba & 
       300 &            &       389 &       438 &       438 & 
                 1.0 &                   5&     0.700 &      0.806 &      0.806\\
$^{151}_{\phantom{1}56}$Ba & 
           &            &       259 &       310 &       310 & 
                     &                    &     1.757 &      4.174 &      4.174\\
$^{152}_{\phantom{1}56}$Ba & 
           &            &       228 &       205 &       205 & 
                     &                    &     2.763 &      4.534 &      4.534\\
$^{153}_{\phantom{1}56}$Ba & 
           &            &       158 &        69 &        69 & 
                     &                    &     4.732 &      4.634 &      4.634\\
$^{154}_{\phantom{1}56}$Ba & 
           &            &       157 &        94 &        94 & 
                     &                    &     6.361 &      9.117 &      9.117\\
\hline
\end{tabular}\\[3ex]
\end{center}
\end{table}
\setcounter{table}{0}
\begin{table}
\caption[tabmas]{
Continued}
\begin{center}
\begin{tabular}{r|rrrrr|rrrrr}
\hline
& \multicolumn{5}{|c|}{$T_{1/2}$ (ms)}
&\multicolumn{5}{c}{$P_{\rm n}$  (\%)} \\
\hline
Isotope & \multicolumn{2}{|c}{Exp.} & KHF & QRPA-1 & QRPA-2
& \multicolumn{2}{|c}{Exp.} & KHF & QRPA-1 & QRPA-2\\
\hline
$^{161}_{\phantom{1}61}$Pm & 
           &            &      1065 &       331 &       331 & 
                     &                    &     0.803 &      0.361 &      0.361\\
 & & & & &   & & & & & \\[-0.12in]
$^{160}_{\phantom{1}62}$Sm & 
      9600 &        300 &      9440 &     26147 &     26147 & 
                     &                    &     0.000 &      0.000 &      0.000\\
$^{161}_{\phantom{1}62}$Sm & 
      4800 &            &      4442 &     11207 &     11207 & 
                     &                    &     0.000 &      0.000 &      0.000\\
$^{162}_{\phantom{1}62}$Sm & 
           &            &      3099 &      6821 &      6821 & 
                     &                    &     0.000 &      0.000 &      0.000\\
$^{163}_{\phantom{1}62}$Sm & 
           &            &      1748 &      3580 &      3580 & 
                     &                    &     0.000 &      0.000 &      0.000\\
\hline
\end{tabular}\\[3ex]
\end{center}
\end{table}

\begin{center}
EXPERIMENTAL DATA
\end{center}

Most of the new $\beta$-decay half-lives of the very neutron-rich 
delayed-neutron precursor isotopes included in Table~\ref{table1} have been 
determined from growth-and-decay curves of neutrons detected with standard 
neutron-longcounter set-ups. As an example, the presently used Mainz 4$\pi$ 
neutron detector consists of 64 $^3$He proportional counters arranged in three 
concentric rings in a large, well-shielded paraffin matrix (B\"ohmer, 1998)
with a total efficiency of about 
45 $\%$. The majority of the new $P_{\rm n}$ values were deduced from the ratios of 
simultaneously measured $\beta$- and delayed-neutron activities. It was only in a few 
cases that $\gamma$-spectroscopic data were used to determine the one or other 
decay property (e.g. independent $P_{\rm n}$ determinations for $^{93}$Br, 
$^{100}$Rb and $^{135}$Sn). Most of the new data were obtained at the on-line 
mass-separator facility ISOLDE at CERN (see, e.g. Fedoseyev {\it et al.}, 1995;
Kratz {\it et al.}, 2000; Hannawald {\it et al.}, 2000; K\"oster, 2000; 
Shergur {\it et al.}, 2000).
Data in the Fe-group region were obtained at the fragment separators LISE at 
GANIL (D\"orfler {\it et al.}, 1996; Sorlin {\it et al.}, 2000)
and FRS at GSI (Ameil {\it et al.}, 1998; Bernas {\it et al.}, 1998),
and at the LISOL separator at Louvain-la-Neuve (Franchoo {\it et al.}, 1998; 
Weissman {\it et al.}, 1999; Mueller {\it et al.}, 2000).
Data in the refractory-element region were measured at the ion-guide separator 
IGISOL at Jyv\"askyl\"a (Mehren {\it et al.}, 1996; Wang {\it et al.}, 1999). 
Finally, 
some new data in the $^{132}$Sn region came from the OSIRIS mass-separator 
group at Studsvik (Korgul {\it et al.}, 2000; Mach {\it et al.}, 2000). \\

In a number of cases, ``old'' $P_{\rm n}$ values from the 1970's deduced from 
measured delayed-neutron yields and (questionable) fission yields not yet 
containing the later well established odd-even effects, were -- as far as 
possible -- corrected, as was also done by Rudstam in his 1993 compilation 
(Rudstam, 1993).
In those cases, where later publications explicitly stated 
that the new data supersede earlier ones, the latter were no longer taken 
into account. Multiple determinations of the same isotopes performed with 
the same method at the same facility by the same authors (e.g. for Rb and Cs 
precursors) were treated differently from the common practice to calculate 
weighted averages of experimental values, when a later measurement
was more reliable than earlier ones. Finally, a number of 
``questionable'' $P_{\rm n}$ values, in particular those where no modern mass model 
would predict the ($Q_{\beta}$ - $S_{\rm n}$) window for neutron emission to be 
positive (e.g. $^{146,147}$Ba and $^{146}$La), are still cited in our Table, 
but should in fact be neglected in any application, hence also in reactor 
calculations.

\begin{center}
MODELS
\end{center}

Theoretically, both integral $\beta$-decay quantities, $T_{1/2}$ and 
$P_{\rm n}$, are interrelated via their usual definition in terms of the 
so-called $\beta$-strength function ($S_{\beta}(E)$) (see, e.g. Duke {\it et 
al.} (1970)).

\begin{equation} \label{thalf}
1/T_{1/2} = \sum_{E_i \geq 0}^{E_i\leq {Q_\beta}}S_\beta(E_i) \times 
f(Z,Q_{\beta}-E_i); 
\end{equation}

\noindent
where $Q_{\beta}$ is the maximum $\beta$-decay energy (or the isobaric mass 
difference) and $f(Z,Q_{\beta}-E_i)$ the Fermi function. With this definition, 
$T_{1/2}$ may yield information on the {\it average} ${\beta}$-feeding of a 
nucleus. However, since the low-energy part of its excitation spectrum is 
strongly weighted by the energy factor of ${\beta}$-decay, 
$f \sim (Q_{\beta}-E_i)^5$, $T_{1/2}$ is dominated by the 
lowest-energy resonances in $S_{\beta}(E_i)$; i.e. by the (near-) ground-state 
allowed (Gamow-Teller, GT) or first-forbidden (ff) transitions. \\

The $\beta$-delayed neutron emission probability ($P_{\rm n}$) is 
schematically given by 

\begin{equation} \label{pn}
P_{\rm n} = \frac{\sum_{B_{\rm n}}^{Q_\beta}S_\beta(E_i)f(Z,Q_\beta-E_i)} 
{\sum_{0}^{Q_\beta}S_\beta(E_i)f(Z,Q_\beta-E_i)}
\end{equation}

\noindent
thus defining $P_{\rm n}$ as the ratio of the integral ${\beta}$-strength to 
states above the neutron separation energy $S_{\rm n}$. As done in nearly all 
$P_{\rm n}$ calculations, in the above equation, the ratio of the partial 
widths for l-wave neutron emission ($\Gamma_{\rm n}^j$($E_{\rm n}$)) and the total width 
($\Gamma_{\rm tot} = \Gamma_{\rm n}^j(E_{\rm n})+\Gamma_{\gamma}$) is set equal to 1; 
i.e. possible $\gamma$-decay from neutron-unbound levels is neglected. As we 
will discuss later, this simplification is justified in most but not all 
delayed-neutron decay (precursor -- emitter -- final nucleus) systems. In any 
case, again because of the ($Q_{\beta}-E)^5$ dependence of the Fermi 
function, the physical significance of the $P_{\rm n}$ quantity is limited, 
too. It mainly reflects the $\beta$-feeding to the energy region just beyond 
$S_{\rm n}$. Taken together, however, the two gross decay properties, 
$T_{1/2}$ {\bf and} $P_{\rm n}$, may well provide some first information 
about the nuclear structure determining $\beta$-decay. Generally speaking, for 
a given $Q_{\beta}$ value a {\it short} half-life usually correlates with  a 
{\it small} $P_{\rm n}$ value, and vice versa. This is actually more
that a rule of thumb since it can be used to 
check the consistency of experimental numbers. Sometimes even global 
plots of double-ratios of experimental to theoretical $P_{\rm n}$ to $T_{1/2}$ 
relations are used to  show  systematic trends (see, e.g. Tachibana {\it 
et al.} (1998)). 
Concerning the 
identification of special nuclear-structure features only from $T_{1/2}$ and 
$P_{\rm n}$, there are several impressive examples in literature. Among them are: 
(i) the development of single-particle (SP) structures and related 
ground-state shape changes in the $50\le N \le 60$ region of the Sr isotopes 
(Kratz, 1984),
(ii) the at that time totally unexpected 
prediction of collectivity of neutron-magic (N=28) $^{44}$S situated two 
proton-pairs below the doubly-magic $^{48}$Ca (Sorlin {\it et al.}, 1993),
and (iii) the very recent interpretation of the surprising 
decay properties of  $^{131,132}$Cd just above $N=82$ (Kratz {\it et al.}, 2000; 
Hannawald {\it et al.}, 2000).\\ 

Today, in studies of nuclear-structure features, even of gross properties 
such as the $T_{1/2}$ and $P_{\rm n}$ values considered here, a substantial 
number of different theoretical approaches are used. The significance and 
sophistication of these models and their relation to each other should, 
however, be clear before they are applied. Therefore, in the following we 
assign the nuclear models used to calculate the above two decay 
properties to different groups:
\begin{enumerate}

\item
{\it Models where the physical quantity of interest is given by an expression 
such as a polynomial  or an algebraic expression}. \\
Normally, the parameters are determined by adjustments to experimental data 
and describe only a single nuclear property. No nuclear wave functions are 
obtained in these models. Examples of theories of this type are purely 
empirical approaches that assume a specific shape of $S_{\beta}$(E) (either
constant or proportional to level density), such as the Kratz-Hermann formula 
(Kratz and Herrmann, 1973) or the statistical ''gross theory'' of $\beta$-decay 
(Takahashi, 1972; Takahashi {\it et al.}, 1973).
These models can be considered to be analogous to the liquid-drop model of 
nuclear masses, and are ---again--- appropriate for dealing with {\it average} 
properties of $\beta$-decay, however taking into account the Ikeda sum-rule to 
quantitatively define the total strength. In both types of  approaches, 
model-inherently no insight into the underlying single-particle (SP) 
structure is possible. \\
 
\item
{\it Models that use an effective nuclear interaction and usually solve
the microscopic 
quantum-mechanical Schr\"odinger or Dirac equation}.  \\
The approaches that actually solve the Schr\"odinger equation
 provide nuclear wave functions which allow a variety of nuclear 
properties (e.g. ground-state shapes, level energies, spins and parities, 
transition rates, $T_{1/2}$, $P_{xn}$, etc.) to be modeled within a single 
framework. Most theories of this type that are currently used in large-scale 
calculations, such as e.g. the FRDM+QRPA model used here 
(M\"oller {\it et al.}, 1997) or the 
ETFSI+cQRPA approach 
(Aboussir {\it et al.}, 1995; Borzov {\it et al.}, 1996), in principle fall 
into two subgroups, 
depending on  the type of microscopic 
interaction used. Another aspect of these models is, whether they are restricted 
to spherical shapes, or to even-even isotopes, or whether they can describe 
{\bf all} nuclear shapes and {\bf all} types of nuclei:
\begin{enumerate}
\item
SP approaches that use a simple central potential with 
additional residual interactions. The Schr\"odinger equation is 
solved in a SP approximation and additional two-body interactions are 
treated in the BCS, Lipkin-Nogami, or RPA approximations, for example. To 
obtain the nuclear potential energy as a function of shape, one combines the 
SP model with a macroscopic model, which then leads to the 
macroscopic-microscopic model. Within this approach, the nuclear ground-state 
energy is calculated as a sum of a microscopic correction obtained from the 
SP levels by use of the Strutinsky method and a macroscopic energy.\\
\item
Hartree-Fock-type models, in which the postulated effective interaction is 
of a two-body type. 
If the microscopic Schr\"odinger equation is solved then
 the wave functions obtained are  antisymmetrized Slater 
determinants. In such models, it is possible to obtain the nuclear ground-state 
energy as \mbox{$E=<\Psi_0|H|\Psi_0>$}, 
otherwise the HF have many similarities to those
in category 2a but have fewer parameters.\\
\end{enumerate}
\end{enumerate}
In principle, models in group 2b are expected to be more accurate, 
because the wave functions and effective interactions can in principle be more 
realistic. However, two problems still remain  today: what effective 
interaction is sufficiently 
realistic to yield more accurate results, and what are the optimized parameter 
values for such a two-body interaction?\\ 

Some models in category 2 have been overparameterized,
which means that their microscopic origins have been lost
and the results are just paramerizations of the experimental data.
Examples of such models are the calculations of Hirsch {\it et al.} (1992, 1996) 
where the strength of the residual GT interaction has 
been fitted for each element (Z-number) in order to obtain optimum 
reproduction of known $T_{1/2}$ and $P_{\rm n}$ values in each isotopic chain. \\
 
To conclude this section, let us emphasize that there is no ``correct'' model 
in nuclear physics. Any modeling of nuclear-structure properties involves 
approximations of the true forces and equations with the goal to obtain a 
formulation that can be solved in practice, but that ``retains the essential 
features'' of the true system under study, so that one can still learn 
something. What we mean by this, depends on the actual circumstances. It may 
well turn out that when proceeding from a simplistic, macroscopic approach to 
a more microscopic model, the first overall result may be ``worse'' just in 
terms of agreement between calcujlated and measured data. 
However, the disagreements  may now be 
understood more easily, and further nuclear-structure-based, realistic 
improvements will become possible.\\

\begin{center} 
PREDICTION OF $P_{\rm n}$ AND $T_{1/2}$ VALUES FROM KHF
\end{center}

As outlined above, Kratz and Herrmann in 1972 (Kratz and Herrmann, 1973) 
applied the concept of the $\beta$-strength function to 
the integral quantity of the delayed-neutron emission probability,  
and derived a simple phenomenological expression 
for $P_{\rm n}$ values, later commonly referred to as the 
{\it ''Kratz-Hermann Formula''} 

\begin{equation} \label{khf}
P_{\rm n} \simeq a [(Q_{\beta}-S_{\rm n})/(Q_{\beta}-C)]^b\qquad [\%] 
\end{equation}

\noindent
where $a$ and $b$ are free parameters to be determined by a log-log fit, and $C$ is 
the cut-off parameter (corresponding to the pairing-gap according to the even 
and odd character of the $\beta$-decay daughter, i.e. the neutron-emitter 
nucleus).\\ 

This KHF has been used in evaluations and in generation of data files (e.g. the ENDF/B 
versions) for nuclear applications up to present. The above free 
parameters $a$ and $b$ were from time to time redetermined 
(Mann {\it et al.}, 1984; Mann, 1986; England {\it et al.}, 1986) as more 
experimental data became available. These values are 
summarized in Table~\ref{oldfits}. Using the present data set presented in 
Table~\ref{table1}, we 
now again obtain new $a$ and $b$ parameters from (i) a linear regression, and (ii) a 
weighted non-linear least-squares fit to about 110 measured $P_{\rm n}$ values in 
the fission-product region. For the present fits, the mass excesses to calculate
$Q_{\beta}$ and $S_{\rm n}$ were taken from the compilation of 
Audi and Wapstra (1995),
otherwise from the FRDM model predictions (M\"oller {\it et al.}, 1995).
The cut-off parameter $C$ was calculated 
according to the expressions given by of Madland and Nix (1988).
With the considerably larger database available today,
apart from global fits of the whole $27\le Z\le 57$ fission-product region, 
also separate fits of the light and heavy mass regions may for the first time be of some 
significance. The corresponding fits to the experimental 
$P_{\rm n}$ values in the different mass regions are shown in Figs.~1--3, and 
the resulting values of the  quantities $a$ and $b$ are given in 
Table~\ref{newfits}. It 
is quite evident from both the Figures and the Tables, that the new fit 
parameters differ significantly from the earlier ones; however, no clear 
trend with the increasing number of experimental data over the years is 
visible. With respect to the present fits, one can state that -- within the 
given uncertainties -- parameter     $a$ does not change very much, neither as a 
function of mass region, nor between the linear regression 
and the non-linear least-squares fit. However, for the slope-parameter $b$ there is a difference.
Here, the least-squares fit consistently results 
in a somewhat steeper slope (by about one unit) than does the linear regression. \\

\begin{figure}[tb]
\vspace{1cm}
\centerline{\psfig{file=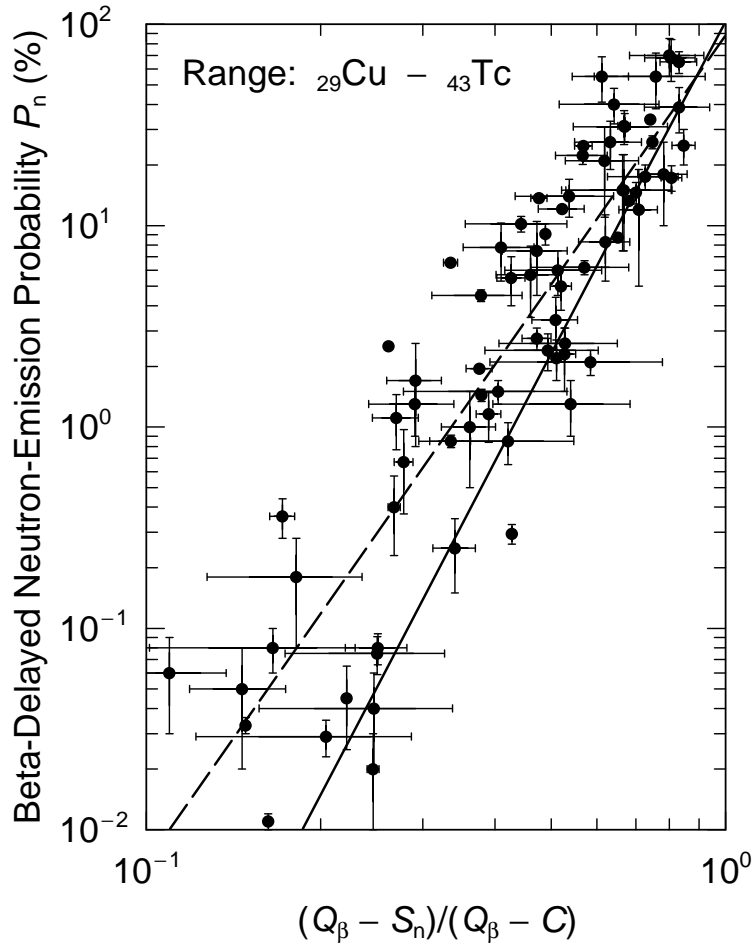,width=10cm}}
\caption[fig1]{Fits to the Kratz-Herrmann-Formula in the region of ``light''
fission products. The measured $P_{\rm n}$ values (dots) 
are displayed as functions of
the reduced energy window for delayed neutron emission. The dotted line is 
derived from a linear regression,
whereas the full line is obtained by a weighted non--linear
least--squares procedure. For the fit parameters, see 
Table~\ref{newfits}.}
\label{fig1}
\end{figure}

\begin{table}[tb]
\caption{Parameters from fits to the Kratz--Herrmann--Formula from 
literature. The two sets from Kratz and Herrmann (1973) derive from different
atomic mass evaluations.}
\begin{center}
\begin{tabular}{l|rr}
\hline
Reference & \multicolumn{2}{c}{Parameters} \\
\hline
 &  $a$ [\%] &  $b$   \\
\hline
Kratz and Herrmann (1973) & 25. & 2.1 $\pm$0.2 \\
\hline
Kratz and Herrmann (1973) & 51. & 3.6 $\pm$0.3  \\
\hline
Mann (1984) & 123.4 & 4.34  \\
\hline
Mann (1986) & 54.0 +31/-20 & 3.44 $\pm$0.51  \\
\hline
England (1986) & 44.08 & 4.119  \\
\hline
\end{tabular}
\end{center}
\label{oldfits}
\end{table}

\begin{figure}[tb]
\vspace{1cm}
\centerline{\psfig{file=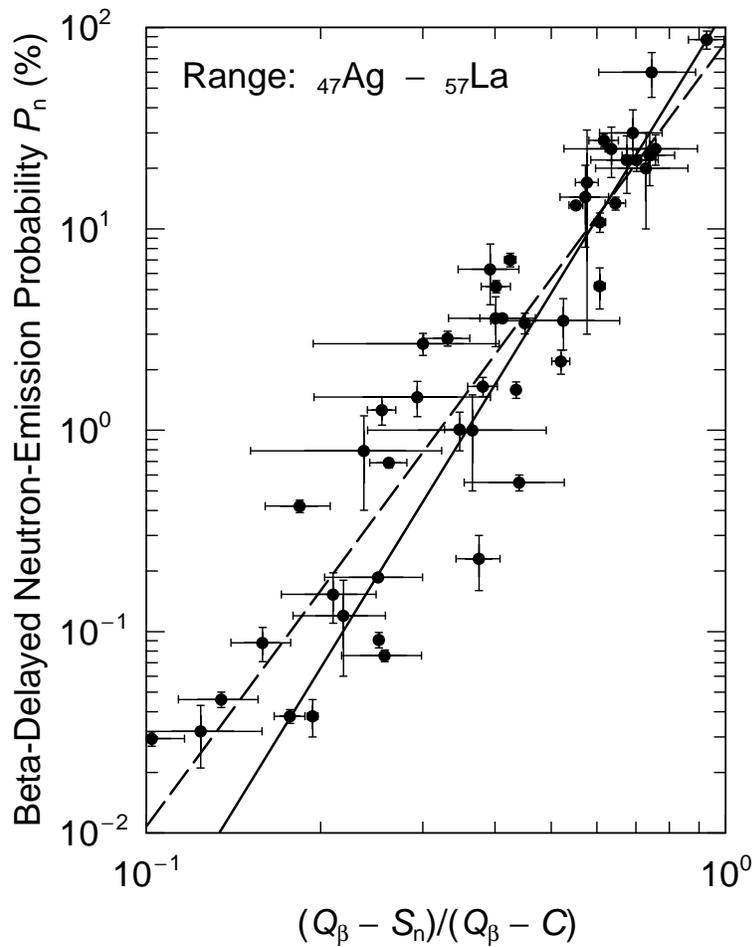,width=10cm}}
\caption[fig2]{Fits to the Kratz-Herrmann-Formula in the region of ``heavy''
fission products. For an explanation of symbols, see Fig.~\ref{fig1}.}
\label{fig2}
\end{figure}

\begin{table}[tb]
\caption{Parameters from fits to the Kratz--Herrmann--Formula in different
mass regions. The sequence corresponds to Figs.~1 to 3.}
\begin{center}
\begin{tabular}{r|rrr|rrr}
\hline
Region & \multicolumn{3}{c}{Lin. regression} & 
\multicolumn{3}{c}{Least-squares fit} \\
\hline
 &  $a$ [\%] &  $b$  & r$^2$ & $a$ [\%] &  $b$ & red. $\chi^2$ \\
\hline
$29\leq Z \leq 43$ & 88.23  &  4.11  &0.81&  105.76  &  5.51 &80.97\\
                  &        &       & & $\pm$37.67 & $\pm$0.61 & \\
\hline
$47\le Z \le 57$ &  84.35  &  3.89  &0.86&  123.09   &   4.68 &57.49 \\
                &         &       & & $\pm$41.17  & $\pm$0.38& \\
\hline
$29\le Z \le 57$ &  85.16  &  3.99 &0.83&  80.58   &   4.72 &78.23 \\
                &         &      & & $\pm$20.72 &  $\pm$0.34 &\\
\hline
\end{tabular}
\end{center}
\label{newfits}
\end{table}

\begin{figure}[tb]
\centerline{\psfig{file=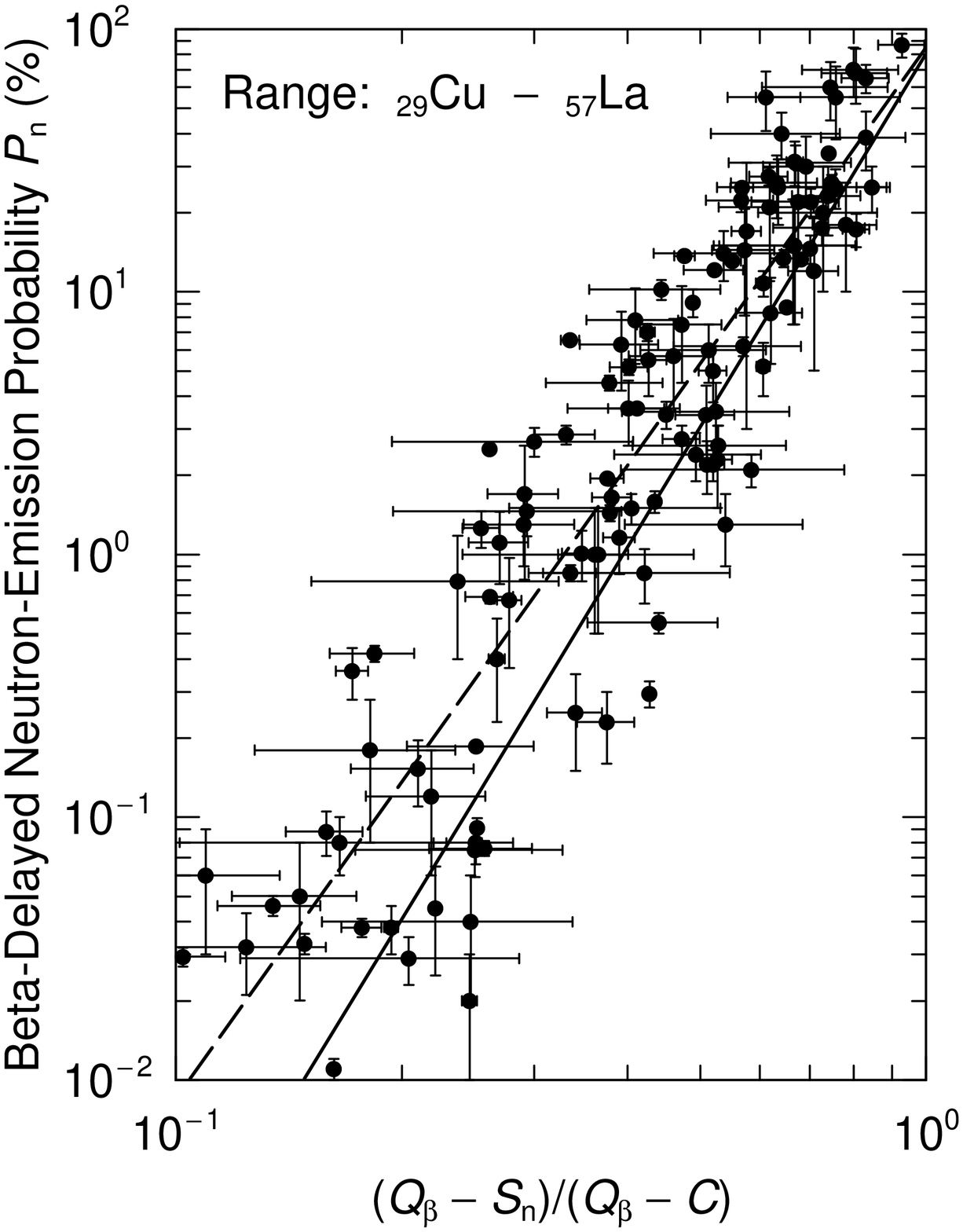,width=10cm}}
\caption{Fits to the Kratz-Herrmann-Formula for all fission products. 
For an explanation of symbols, see Fig.~\ref{fig1}.}
\label{fig3}
\end{figure}

Based on the new non-linear least-squares fit parameters, the KHF was used to 
predict so far unknown $P_{\rm n}$ values between $_{27}$Co and $_{63}$Eu in the 
relevant mass ranges for each isotopic chain.
These theoretical values are listed in  Table~\ref{table1}. \\

In analogy with  the $P_{\rm n}$ values, the $\beta$-decay half-lives $T_{1/2}$ are to be regarded
as ``gross'' properties. Therefore, one can assume that the statistical concepts
underlying the Kratz--Herrmann-formula for $P_n$ values can be applied for the
description of $T_{1/2}$.\\
The half-lives are inversely proportional to the Fermi-function $f(Z,E)$, which,
in first order, is proportional to the fifth power of the reaction $Q_{\beta}$-value:
\begin{equation} \label{eqthalfkhf}
T_{1/2} \sim 1/f(Z,E) \sim Q_\beta^{-5}
\end{equation} 
Therefore, in a log-log plot of $T_{1/2}$ versus $Q_\beta$ one expects the data
points to be scattered around a line with a slope of about -(1/5). 

Pfeiffer {\it et al.} (2000) suggested to fit the $T_{1/2}$ of 
neutron-rich nuclides according to the following expression:

\begin{equation} \label{t1/2}
T_{1/2} \simeq a \times \left( Q_{\beta} - C  \right) ^{-b}
\end{equation}
where the cut-off parameter $C$ is 
calculated according to the fit of Madland and Nix (1988),
and the parameters $a$ and $b$ are listed in Table~\ref{t12}.

The gross theory has, basically, the same functional dependence on the 
$Q_{\beta}$-value, but 
underestimates the $\beta$-strength to low-lying states, which results in too long
half-lives.  We here  compensate for this deficiency by treating the coefficient $a$ 
as a free parameter to be determined by a fitting procedure.
The  values obtained are listed in Table~\ref{table1}.
\large{
\begin{table}[tb]
\caption{Parameters from fits to $T_{1/2}$ of neutron--rich nuclides.}
\label{t12}
\begin{center}
\begin{tabular}{rrr|rrr}
\hline
  \multicolumn{3}{c}{lin. regression} & 
\multicolumn{3}{c}{least-squares fit} \\
\hline
 $a$ [ms] &  $b$  & r$^2$ & $a$ [ms] &  $b$ & red. $\chi^2$ \\
\hline
 2.74E06 & 4.5 &0.72 & 7.07E05  &  4.0 &1.1E04 \\
                     &      & & $\pm$5.33E05 & $\pm$0.4 & \\
\hline
\end{tabular}
\end{center}
\end{table}
}

\begin{center}
PREDICTION OF $T_{1/2}$ AND $P_{\rm n}$ VALUES FROM FRDM-QRPA
\end{center}

The formalism we use to calculate Gamow-Teller (GT) $\beta$-strength
functions is fairly lengthy, since it involves
adding  pairing and Gamow-Teller residual interactions
to the folded-Yukawa single-particle Hamiltonian and
solving the resulting Schr\"{o}dinger equation in the
quasi-particle random-phase approximation (QRPA). Because
this model has been completely described
in two previous papers (Krumlinde {\it et al.\/}, 1984; M\"oller 
{\it et al.\/}, 1990),
we refer to those two publications for a full
model specification and for a definition of notation used.
We restrict the discussion here to an overview of features that
are particularly relevant to the results discussed in
this paper.

It is well known that wave functions and transition
matrix elements are more affected
by small perturbations to the Hamiltonian than are the
eigenvalues. When transition rates are
calculated it is therefore necessary to add
residual interactions to the folded-Yukawa single-particle
Hamiltonian in addition to the pairing interaction that
is included in the mass model. Fortunately,
the residual interaction may be restricted
to a term specific to the particular
type of decay considered. To obtain reasonably accurate half-lives
it is also very important to include ground-state deformations. Originally the
QRPA formalism was developed for and applied only to spherical
nuclei (Hamamoto, 1965; Halbleib {\it et al.\/}, 1967). The extension to
deformed nuclei, which is necessary in global calculations of
$\beta$-decay properties, was first described in 1984 (Krumlinde 
{\it et al.\/}, 1984).
 
To treat Gamow-Teller $\beta$ decay  we therefore add
the Gamow-Teller  force
\beq
V_{\rm GT}=2\chi_{\rm GT}:\mbox{\boldmath $\beta^{1-}\cdot\beta^{1+}$}\!:
\eeq{vgt}
to the folded-Yukawa single-particle Hamiltonian,
after pairing has already been
incorporated,
with the standard choice $\chi_{\rm GT}=
23$~MeV/$A$ (Hamamoto, 1965; Halbleib {\it et al.\/}, 1967;
Krumlinde {\it et al.\/}, 1984; M\"oller {\it et al.\/}, 1990).
Here \mbox{\boldmath $\beta^{1\pm}$}$= \sum_i$\mbox{\boldmath
$\sigma_it_i^{\pm}$} are the Gamow-Teller $\beta^{\pm}$-transition operators.

The process of $\beta$ decay occurs from an initial ground state or excited
state in
a mother nucleus to a final state in the daughter nucleus.
For $\beta^-$ decay, the final configuration
is a nucleus in some excited state or its ground state,
an electron (with energy $E_{\rm e}$), and an anti-neutrino (with energy
$E_{\nu}$).
The decay rate $w_{fi}$ to one nuclear state $f$ is
\beq
w_{fi} = \frac{m_0c^2}{\hbar} \;
\frac{\Gamma^2}{2\pi^3}
\;|M_{fi}|^2
f(Z,R,\epsilon_0)
\eeq{distint}
where $R$ is the nuclear radius and
$\epsilon_0=E_0/m_0c^2$, with
$m_0$  the electron mass.
Moreover, $|M_{fi}|^2$ is the nuclear matrix element,
which is also the $\beta$-strength function.
The dimensionless constant $\Gamma$ is defined by
\beq
\Gamma \equiv \frac{g}{m_0c^2} {\left( \frac{m_0c}{\hbar} \right)}^3
\eeq{gamma}
where $g$ is the Gamow-Teller coupling constant.
The quantity $f(Z,R,\epsilon_0)$ has been extensively discussed
and tabulated elsewhere (Preston, 1962; Gove and Martin, 1971; deShalit
and Feshbach, 1974).
 
For the special case in which  
the two-neutron separation energy $S_{\rm 2n}$ in the daughter nucleus
is greater
than  $Q_{\beta}$, the energy released in ground-state to ground-state
  $\beta$ decay,
the probability for $\beta$-delayed one-neutron emission,
in percent, is given by
\beq
P_{\rm 1n} = 100 \, \frac
{\begin{displaystyle}\sum_{S_{\rm 1n}<E_f<Q_{\beta}} w_{fi}\end{displaystyle}}
{\begin{displaystyle}\sum_{0<E_f<Q_{\beta}} w_{fi}\end{displaystyle}}
\eeq{deln}
where 
$E_{f}=Q_{\beta} -E_0$ is the excitation energy in the daughter nucleus 
and $S_{\rm 1n}$ is the one-neutron
separation energy in the daughter nucleus.
We assume that decays to energies above
$S_{\rm 1n}$ always lead to delayed neutron emission.
 
To obtain the half-life with respect to $\beta$ decay
one sums up the decay rates $w_{fi}$ to the individual nuclear
states in the allowed energy window. The half-life is then related to the total
decay rate by
\beq
T_{\beta} = \frac{\ln 2}{\begin{displaystyle}\sum_{0<E_f<Q_{\beta}}
 w_{fi}\end{displaystyle}}
\eeq{beha}
The above equation may be rewritten as
\beq
T_{\beta}  =  \frac{\hbar}{m_0c^2} \; \frac{2\pi^3\ln 2}{\Gamma^2}
\frac{1}{\begin{displaystyle}\sum_{0<E_f<Q_{\beta}} \;|M_{fi}|^2
f(Z,R,\epsilon_0)
\end{displaystyle}}
 =  \frac{B}{\begin{displaystyle}\sum_{0<E_f<Q_{\beta}} \;|M_{fi}|^2
f(Z,R,\epsilon_0)
\end{displaystyle}}
\eeq{beha2}
with
\beq
B  = \frac{\hbar}{m_0c^2} \; \frac{2\pi^3\ln 2}{\Gamma^2}
\eeq
{beha3}
For the value of $B$ corresponding to Gamow-Teller decay
we use
\beq
B= 4131 \; {\rm s}
\eeq{bvalue}
 
The energy released in ground-state to ground-state electron
decay is given  in terms of the atomic mass excess
 $M(Z,N)$ or the total binding
energy  $E_{\rm bind}(Z,N)$ by
\beq
Q_{\beta^-}= M(Z,N) - M(Z+1,N-1) 
\eeq{erelel}

The above formulas apply to the $\beta^-$ decays that are of interest here.
The decay $Q$ values and neutron separation energies $S_{\nu\rm n}$
are obtained from our FRDM  mass model when experimental data are 
unavailable (M\"oller {\it et al.}, 1995).  The matrix elements $M_{fi}$ are 
obtained
from our QRPA model. More details are provided elsewhere (M\"oller {\it et
al.\/}, 1990).

We present here two calculations, QRPA-1 and QRPA-2 of $T_{1/2}$ and 
$P_{\rm n}$. 
\begin{figure}[t]
\centerline{\psfig{file=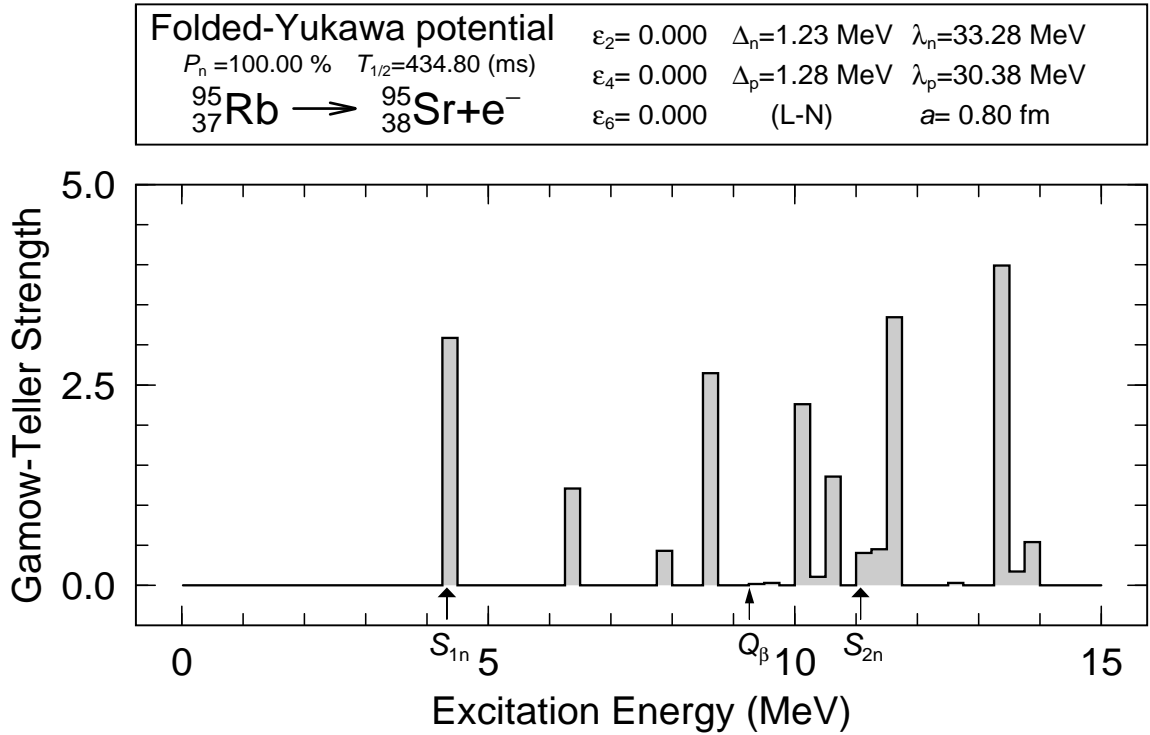,width=6in}}
\caption{Calculated $\beta$-strength function for $^{95}$Rb in
our standard model (M\"oller {\it et al.\/}, 1997). However, the deformation
is not taken from the standard ground-state mass and deformation calculation
(M\"oller {\it et al.\/}, 1995). Instead the ground-state shape is assumed spherical,
in accordance with experimental evidence. The figure shows the sensitivity
of the calculated $P_{\rm n}$ value to small details of the model. Since
there is no strength below the neutron separation energy, the calculated
$\beta$-delayed neutron-emission probability is 100\%. However
it is clear from the figure that just a small decrease in the energy
of the large peak just above the neutron binding energy would
drastically change the calculated value.}
\label{rb95a}
\end{figure}
\begin{figure}[t]
\centerline{\psfig{file=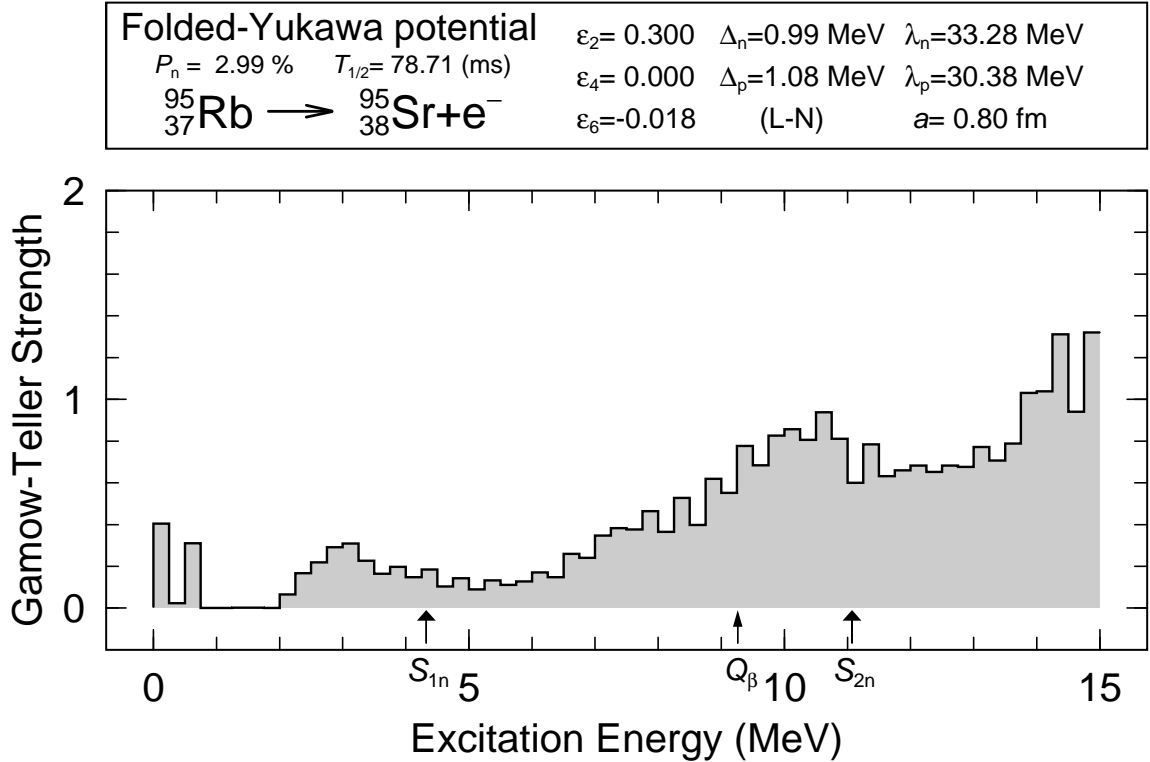,width=6in}}
\caption{This calculation corresponds to the {\bf QRPA-1}
model specification. However, this nucleus is known to be spherical
although  a deformed shape was obtained in
the ground-state mass-and-deformation calculation (M\"oller {\it et al.\/}, 1995). 
Therefore, in our {\bf QRPA-2} calculation
in Fig.~\ref{rb95c}, this nucleus is treated as spherical in
accordance with experiment.}
\label{rb95b}
\end{figure}
\begin{figure}[t]
\centerline{\psfig{file=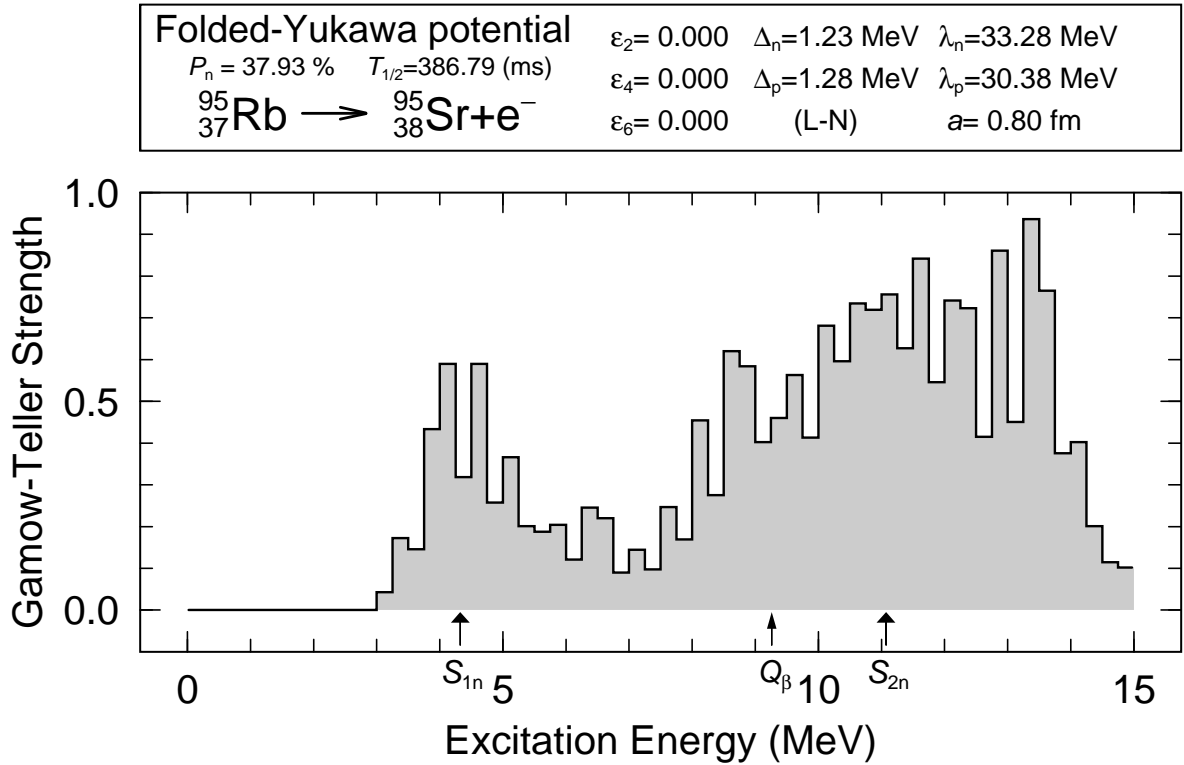,width=6in}}
\caption{This calculation corresponds to the {\bf QRPA-2}
model specification. The calculation is identical to
the calculation in Fig.~\ref{rb95b} except that the ground-state
shape here is spherical.}
\label{rb95c}
\end{figure}
They are based on
our standard $QRPA$ model described above, but with the following enhancements:
\begin{description}
\newpage
\item[For QRPA-1:]\mbox{ }\\[-0.25in]
\begin{enumerate}
\item
To calculate $\beta$-decay $Q$-values and neutron separation energies
$S_{\nu{\rm n}}$ we use experimental ground-state masses where available,
otherwise calculated masses (M\"oller {\it et al.\/}, 1995). In our
previous recent calculations we used the 1989 mass evaluation (Audi 1989);
here we use the 1995 mass evaluation (Audi {\it et al.\/}, 1995).
\item
It is known that at higher excitation energies additional residual interactions
result in a spreading of the transition  strength. In our 1997 calculation
each transition goes to a precise, well-specified energy in the daughter
nucleus. This can result in very large changes in the calculated $P_{\rm n}$
values for minute changes in, for example $S_{\rm 1n}$, depending on whether
an intense, sharp transition is located just below or just above the neutron 
separation energy (M\"oller {\it et al.\/}, 1990). To remove this unphysical 
feature
we introduce an empirical spreading width that sets in above 2 MeV.
Specifically, each transition strength ``spike'' above 2 MeV
is transformed to a Gaussian of width
\beq
\Delta_{\rm sw}=\frac{8.62}{A^{0.57}}
\eeq{spreadw}
This choice is equal to the error in the mass model. Thus, it accounts
approximately for the uncertainty in calculated neutron separation
energies and at the same time it roughly corresponds to the
observed spreading of transition strengths in the energy range
2--10 MeV, which is the range of interest here.
\end{enumerate}
\item[For QRPA-2:]\mbox{ }\\[-0.25in]
\begin{enumerate}
\item
In this calculation we retain all of the features of the QRPA-1 
calculation and in addition  account more accurately for the 
ground-state deformations which affect the energy levels
and wave-functions that are obtained in the single-particle model.
The ground-state deformations calculated in the FRDM mass model 
(M\"oller {\it et al.}, 1992),
generally agree with experimental observations, but in transition
regions between spherical and deformed nuclei discrepancies do occur.
In the QRPA-2 calculation we therefore replace calculated
deformations with spherical shape, when experimental data so indicate. 
This has been done for the following nuclei:\\
$^{67-78}$Fe,
$^{67-79}$Co,
$^{73-80}$Ni,
$^{73-81}$Cu,
$^{78-84}$Zn,
$^{79-87}$Ga,
$^{83-90}$Ge,
$^{84-91}$As,
$^{87-94}$Se,
$^{87-96}$Br,
$^{92-98}$Kr,
$^{91-96}$Rb,
$^{96-97}$Sr,
$^{96-98}$Y,
$^{134-140}$Sb,
$^{136-141}$Te,
$^{137-142}$I,
$^{141-143}$Xe, and
$^{141-145}$Cs.
 
\end{enumerate}
\end{description}

To illustrate some typical features of $\beta$-strength functions we
present the strength function of $^{95}$Rb calculated in three
different ways in Figs.~\ref{rb95a}--\ref{rb95c}. 
 
\begin{figure}[t]
\centerline{\psfig{file=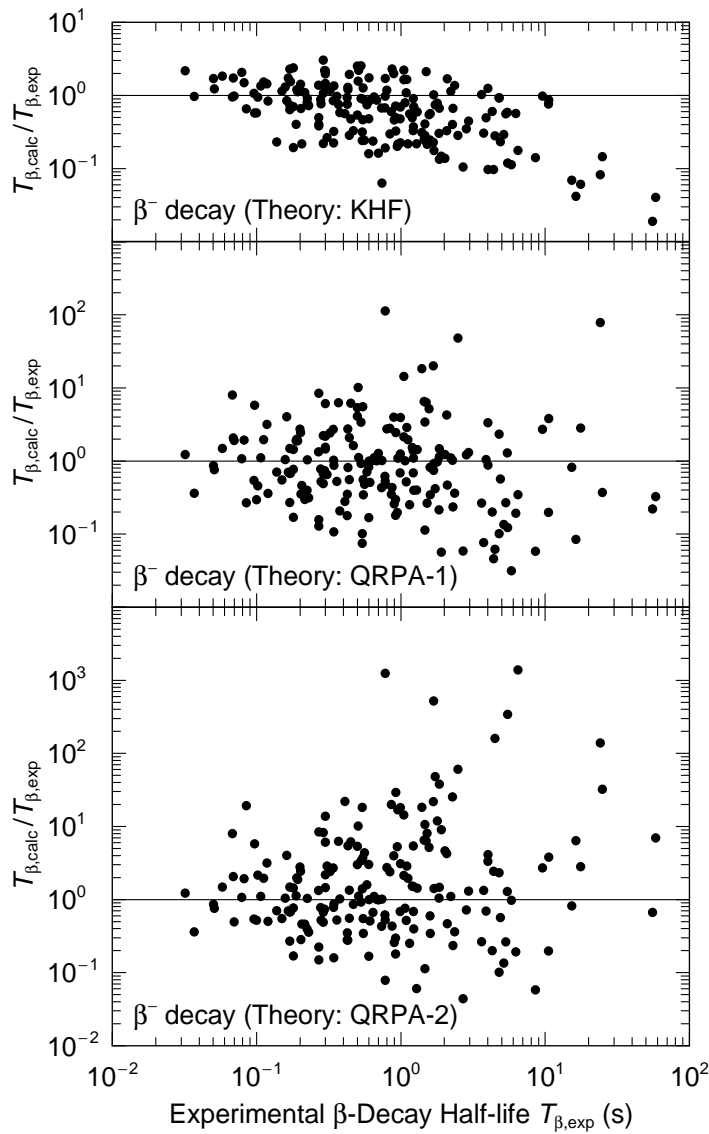,height=6in}}
\caption{Ratio of calculated to experimental $\beta$-decay half-lives
for nuclei in the fission-product region in three different models.}
\label{betlif}
\end{figure}

\begin{figure}[t]
\centerline{\psfig{file=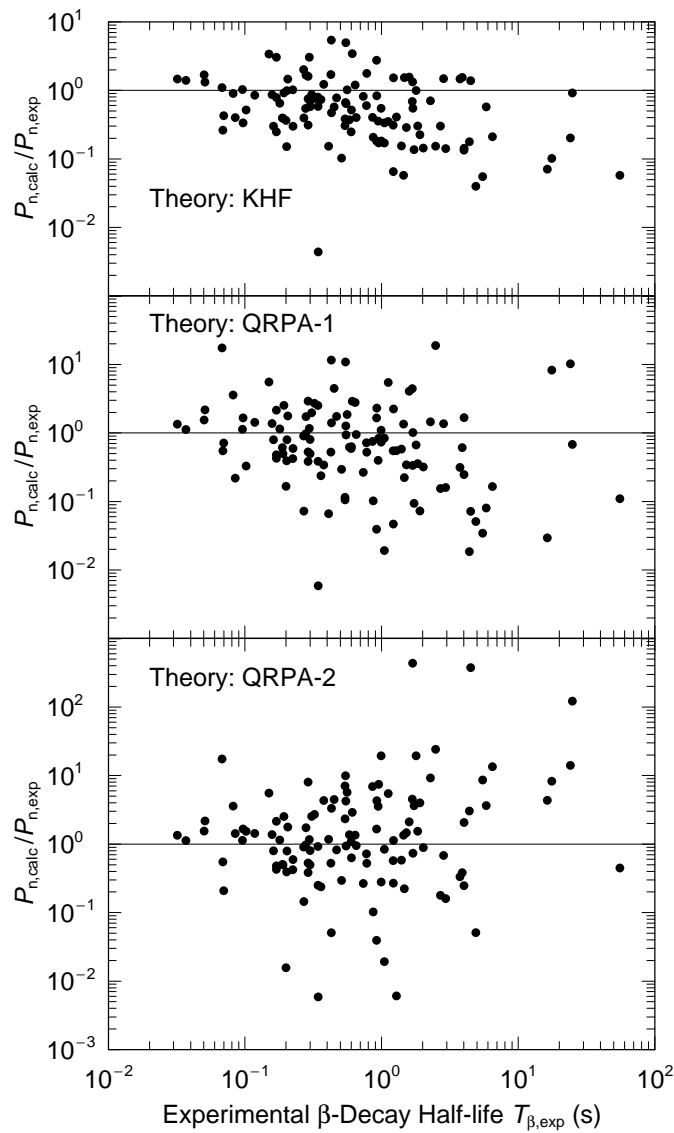,height=6in}}
\caption{Ratio of calculated to experimental $\beta$-delayed
neutron-emission probabilities 
for nuclei in the fission-product region in three different models.}
\label{betpn}
\end{figure}

It is not our aim here to make a detailed analysis of each individual
nucleus, but instead to present an overview of the model performance in
a calculation of a large number of $\beta$-decay half-lives.  In
Figs.~\ref{betlif} and \ref{betpn} we compare measured 
and calculated
$\beta$-decay half-lives and $\beta$-delayed
neutron emission probabilities for the nuclei considered here.
To address the reliability in various regions
of nuclei and versus distance from stability, we present 
the ratios 
$T_{\beta,{\rm calc}}/T_{\beta,{\rm exp}}$ 
$P_{{\rm n},{\rm calc}}/P_{{\rm n},{\rm exp}}$ 
versus the quantity $T_{\beta,{\rm exp}}$.
 Because the relative error in the calculated
half-lives is more sensitive to small shifts in the positions of the
calculated single-particle levels for decays with small energy releases,
where long half-lives are expected, one can anticipate that
half-life calculations are more reliable far from stability than close
to $\beta$-stable nuclei. 
\begin{table}[tb]    
\caption[betlifmt]  
{Analysis of the discrepancy between calculated and \label{tabliferr}
measured $\beta^-$-decay half-lives shown in Fig.~\ref{betlif}.}
\begin{center}      
\begin{tabular}{rrrrrrr}
\hline\\[-0.07in]   
Model & $ n $ & $M_{r_{\rm l}}$ & $ M_{r_{\rm l}}^{10} $ & $  \sigma_{r_{\rm l}}$
 & $\sigma_{r_{\rm l}}^{10}$ & $T_{\beta,{\rm exp}}^{\rm max}$\\
&  & & &  & & (s)\\[0.08in]
\hline\\[-0.07in]   
KHF    &  115&  $-0.11$&   0.77&   0.33&   2.15&     1\\
QRPA-1 &  115&  $-0.02$&   0.95&   0.50&   3.14&     1\\
QRPA-2 &  115&     0.13&   1.37&   0.61&   4.04&     1\\[1ex]
KHF    &  187&  $-0.40$&   0.58&   0.41&   2.56&    all\\
QRPA-1 &  187&  $-0.06$&   0.87&   0.59&   3.88&    all\\
QRPA-2 &  187&     0.22&   1.67&   0.75&   5.75&    all\\[1ex]
\hline              
\end{tabular}       
\end{center}        
\end{table}         
\begin{table}[tb]    
\caption[betpner]  
{Analysis of the discrepancy between calculated and \label{tabpnerr}
measured $\beta$-delayed neutron-emission 
probabilities $P_{\rm n}$ values  shown in Fig.~\ref{betpn}.}
\begin{center}      
\begin{tabular}{rrrrrrr}
\hline\\[-0.07in]   
Model & $ n $ & $M_{r_{\rm l}}$ & $ M_{r_{\rm l}}^{10} $ & $  \sigma_{r_{\rm l}}$
 & $\sigma_{r_{\rm l}}^{10}$ & $P_{\rm n,exp}^{\rm min}$\\
&  & & &  & & (\%)\\[0.08in]
\hline\\[-0.07in]   
KHF    &   86&  $-0.31$&   0.49&   0.36&   2.31&     1\\
QRPA-1 &   86&  $-0.12$&   0.76&   0.60&   4.02&     1\\
QRPA-2 &   86&     0.12&   1.34&   0.65&   4.51&     1\\[1ex]
KHF    &  118&  $-0.29$&   0.51&   0.44&   2.76&    all\\
QRPA-1 &  118&  $-0.18$&   0.66&   0.62&   4.14&    all\\
QRPA-2 &  118&     0.11&   1.28&   0.75&   5.62&    all\\[1ex]
\hline              
\end{tabular}       
\end{center}        
\end{table}

Before we make a quantitative analysis of the agreement between
calculated and experimental half-lives we briefly discuss what
conclusions can be drawn from a simple visual
inspection of Figs.~\ref{betlif} and \ref{betpn}.
As  functions of $T_{\beta,{\rm exp}}$
one would expect the average error to increase as
$T_{\beta,{\rm exp}}$ increases.  This is indeed the case.
In addition one is
left with the impression that the errors in our calculation are fairly
large. However, this is partly a fallacy, since for small errors there
are many more points than for large errors. This is not clearly seen in
the figures, since for small errors many points are superimposed on
one another. To obtain a more exact understanding of the error in the
calculation we therefore perform a more detailed analysis.

One often analyzes the error in a calculation by studying a
root-mean-square deviation, which in this case would be
\beq                
{\sigma_{\rm rms}}^2 = \frac{1}{n}\sum_{i=1}^{n} (T_{\beta,{\rm exp}} -
T_{\beta,{\rm calc}})^2
\eeq{sigrms}        
However, such an error analysis is
unsuitable here, for two reasons.  First, the quantities studied vary
by many orders of magnitude. 
Second, the calculated  and measured quantities may {\it differ} by orders of
magnitude. We therefore study the quantity $\log
(T_{\beta,{\rm calc}}/T_{\beta,{\rm exp}})$, which is plotted  in
Fig.~\ref{betlif},
instead of $(T_{\beta,{\rm exp}} - T_{\beta,{\rm calc}})^2$.
We present the formalism here for the half-life, but the formalism
is also used to study the error of our calculated $P_{\rm n}$ 
values. 
 
To facilitate the interpretation of the error plots we consider two
hypothetical cases.  As the first example, suppose that all the points
were grouped on the line $T_{\beta,{\rm calc}}/T_{\beta,{\rm
exp}}=10$.  It is immediately clear that an error of this type could be
entirely removed by introducing a  renormalization factor, which is a
common practice in the calculation of $\beta$-decay half-lives.  We
shall see below that in our model the half-lives corresponding to our
calculated  strength functions have about zero average deviation from
the calculated half-lives, so no renormalization factor is necessary.
 
In another extreme, suppose half the
points were located on the line
$T_{\beta,{\rm calc}}/T_{\beta,{\rm exp}}=10$
and the other half on the line
$T_{\beta,{\rm calc}}/T_{\beta,{\rm exp}}=0.1$.
In this case the average of
$\log(T_{\beta,{\rm calc}}/T_{\beta,{\rm exp}})$
would be zero.  We are
therefore led to the conclusion that there are two types of errors that
are of interest to study, namely the average position of the points in
Fig.~\ref{betlif},
which is just the average of the quantity
$\log(T_{\beta,{\rm calc}}/T_{\beta,{\rm exp}})$,
and the spread of the points around this
average.  To analyze the error along these ideas, we
introduce the quantities
\beqar              
                 r &=&T_{\beta,{\rm calc}}/T_{\beta,{\rm exp}} \nonumber \\[1ex]
                 r_{\rm l} &=& \log_{10}(r) \nonumber \\[1ex]
                 M_{r_{\rm l}} &=&
                    \frac{1}{n}\sum_{i=1}^n r_{\rm l}^i \nonumber \\[1ex]
                 M_{r_{\rm l}}^{10} &=& 10^{M_{r_{\rm l}}}   \nonumber \\[1ex]
\sigma_{r_{\rm l}} &=& {\left[
\frac{1}{n} \sum_{i=1}^n {\left( r_{\rm l}^i
- M_{r_{\rm l}} \right)}^2 \right] }^{1/2} \nonumber \\[1ex]
\sigma_{r_{\rm l}}^{10} &=& 10^{\sigma_{r_{\rm l}}}
\eeqar{statdef}     
where $M_{r_{\rm l}}$ is the average position of the points
and $\sigma_{r_{\rm l}}$ is the spread around this average.
The spread $\sigma_{r_{\rm l}}$ can be expected to be related
to uncertainties in the positions of the levels in the
underlying single-particle model.
The use of a logarithm in the definition of
$r_{\rm l}$ implies that these two
quantities correspond directly to distances as seen by the eye in
Figs.~\ref{betlif}--\ref{betpn},
in units where one order of magnitude is 1. After the error
analysis has been carried out we want to discuss its result
 in terms like ``on the average the calculated half-lives
are `a factor of two' too long.'' To be able to do this we must convert
back from the logarithmic scale. Thus, we realize
that the quantities $M_{r_{\rm l}}^{10}$ and $\sigma_{r_{\rm l}}^{10}$
are conversions back to ``factor of'' units of
the quantities $M_{r_{\rm l}}$ and $\sigma_{r_{\rm l}}$,
which are expressed in distance
or logarithmic units.\\[1ex]
\begin{center}
DISCUSSION AND SUMMARY
\end{center}
 
In Tables \ref{tabliferr} and \ref{tabpnerr}
we show the results of an evaluation of the quantities in
Eq.~(\ref{statdef}) for $T_{1/2}$ and $P_{\rm n}$ 
corresponding to $\beta$ decay of the nuclei
in table \ref{table1}.
In the QRPA calculations the ratio between calculated
and measured decay half-lives is close to 1.0. This
shows, as pointed out earlier (M\"oller and Randrup, 1990)
that {\it no} renormalization of the calculated strength is
necessary. The mean deviation between calculated and experimental
half-lives is a factor of 2--5 depending on model and half-life
cutoff. Also the  calculated
$P_{\rm n}$  values agree on the average with the experimental data.
Here the mean deviation between calculated and experimental
data is a factor of 3--6, again depending on model and half-life cutoff.
All half-life calculations agree better with data for shorter half-lives,
cf. Fig.~\ref{betlif} and  Table~\ref{tabliferr}.
Therefore one can expect the models to perform better far from stability than 
what is indicated by the table. The $\beta$-delayed neutron emission rates
are also better calculated in the region of short half-lives and
high $P_{\rm n}$ values,
cf. Fig.~\ref{betpn} and  Table~\ref{tabpnerr}. Again, this suggests
calculated $P_{\rm n}$ values are more reliable far from
stability than indicated by Table~\ref{tabpnerr}.

The KHF results appear more reliable than the QRPA results.
This may seem surprising at first, because the KHF has minimal
microscopic content compared to the QRPA. However, 
an advantage of the QRPA is that it
provides so much more detail about $\beta$-decay than does
the KHF, namely the $ft$ values of the individual decays,
and the transition energies associated with those decays.
A very detailed discussion of the possible sources
of discrepancies between our QRPA results and experimental
data is presented in Ref.~(M\"{o}ller and Randrup, 1990).
One difficulty the calculations face is that the
calculated half-lives depend on the energy of
the transitions as $(Q_{\beta}-E)^5$. As an example
we note that calculated half-lives for $^{95}$Rb,
for which $Q_{\beta}=9.28$~MeV,
change by a factor 1.5 for a change in transition energies 
by only 0.4 MeV. It is very difficult to reproduce transition
energies to this accuracy in a global nuclear-structure model.

For the QRPA-2 calculation we observe that the average of
$T_{\beta,{\rm calc}}/T_{\beta,{\rm exp}}$ is considerably larger
than 1, which corresponds to a correct average. One would have
{\it a priori} assumed that this calculation would be in better
agreement with experiment since we substitute {\it calculated}
deformations for spherical deformations when so indicated by
experimental data. However, since we do not include
$\beta$-strength due to forbidden transitions in our
model, one would indeed expect that calculated half-lives
be too low on the average. The non-spherical deformations
that occur, contrary to experimental observations, in
the QRPA-1 calculations in some sense simulate the missing
low-lying forbidden $\beta$-strength. However, a much more satisfying
description would be to use correct ground-state deformations and 
develop some model to account for the strength related to forbidden
transitions.

The $P_{\rm n}$ values calculated in the QRPA-1 are on the
average too low. At present we have no clear explanation for this result.
An obvious correction to the model is to take competition with
$\gamma$ emission into account, in particular for emission of $l_{\rm n}\ge 3$
neutrons. However, such a correction would further lower the ratio
$T_{\rm n,calc}/T_{\rm n,exp}$. One may speculate that an accounting
for {\it both} this effect and forbidden transition strength in
QRPA-2 would bring about satisfactory agreement. This possibility
need to be investigated.

We feel strongly that in a global, unified nuclear-structure model a single set
of constants must be used over the entire chart of the nuclides, otherwise the
basic foundation of the model is violated. However, for the purpose of
generating the best possible data bases of half-lives and $\beta$-delayed
neutron-emission probabilities a complementary approach is reasonable. Just as
we feel it is appropriate to use experimental ground-state deformations,
experimental single-particle levels, when known, could also be used as the
starting point for the QRPA calculations. In practice the situation would be
that in some regions, such as near the doubly magic $^{132}$Sn, many half-lives
and $P_{\rm n}$ values would be unknown, but considerable information on
single-particle level order and energies would be available. This experimental
information could then be taken into account by locally adjusting the
single-particle model proton and neutron spin-orbit strengths and the
diffuseness of the single-particle well to obtain optimum agreement with the
observed single-particle data such as the observed neutron single-particle
sequence $f_{7/2}$, $p_{3/2}$, $p_{1/2}$, and $h_{9/2}$ near $^{132}$Sn.  The
hope would be that the local agreement would be retained in some limited
extrapolation away from the known region. Such a fairly limited extrapolation
would be all that is required to reach the isotopes in the fission-product
region where experimental data are not yet available, cf. Fig.~\ref{chart}.
Limited studies along these lines have been undertaken by, for example,
Hannawald {\it et al.\/} (2000).
Other highly desirable enhancements to the calculations would be to include
first-forbidden strength, perhaps first in a gross-theory approach and later
from a new microscopic model.  The cut-off parameter $C$ in the KHF formula
could be taken from the Lipkin-Nogami microscopic calculation instead of from
the Madland-Nix macroscopic expression.  The energy window ($Q_\beta - S_{\rm
n})$ could be reduced by 150 to max 500 keV to account for the angular-momentum
barrier for emission of $l\geq 3$ neutrons in for example $^{137}$I.

In conclusion we note that we now have available about 40 new experimental
$T_{1/2}$ and $P_{\rm n}$ values in the fission-product region. Data for
additional nuclei in this region that are required as input in reactor
criticality, astrophysical and other applications are provided from theoretical
calculations. The substantial increase in available experimental data since the
compilations by Brady (1989) and Rudstam (1993) is expected to 
have a significant impact on applied calculations.

\begin{center}
REFERENCES
\end{center}

\noindent
Aboussir Y.{\it et al.\/} (1995). {\it Atomic Data and Nucl. Data Tables} 
   {\bf 61}, 127. \\ 
Ameil F.{\it et al.\/} (1998).  {\it Eur. Phys. J.} {\bf A1}, 275. \\
Audi, G. (1989).
``Midstream atomic mass evaluation, private communication,
with four revisions''\\
Audi G. and Wapstra A.H. (1995). {\it Nucl.Phys.} {\bf A595}, 409. \\ 
Bernas M. {\it et al.\/} (1998).  {\it Nucl. Phys.} {\bf A630}, 41c.\\
B\"ohmer W. (1998). PhD Thesis, Univ. Mainz, unpublished. \\
Borzov I.N. {\it et al.\/} (1996). {\it Z. Phys.} {\bf A335}, 127. \\ 
Brady M.C. (1989). ``Evaluation and Application of Delayed Neutron 
    Precursor Data'', LANL Thesis Report LA-11534-T. \\
Brady M.C. and England T.R. (1989). {\it Nucl. Sci. Eng.} {\bf 103}, 129. \\
D\"orfler T. {\it et al.\/} (1996).  {\it Phys. Rev.} {\bf C54}, 2894. \\
Duke C.L., Hansen P.G., Nielsen O.B. and Rudstam G.(1970). 
  {\it Nucl. Phys.} {\bf A151}, 609. \\ 
England T.R. {\it et al.\/} (1986). Specialists' Meeting on Delayed Neutron 
    Properties ISBN 070044 0926 7, p.117. Birmingham, Sep. 1986. \\
Fedoseyev V.N. {\it et al.\/} (1995).  {\it Z. Physik} {\bf A353}, 9. \\
Franchoo S. {\it et al.\/} (1998).  {\it Phys. Rev. Lett.} {\bf 81}, 3100. \\ 
Gove N.B. and Martin M.J. (1971). {\it Nucl.\ Data Tables} {\bf 10}, 205.\\
Halbleib Sr J.A. and Sorensen R. A. (1967). {\it Nucl. Phys.} {\bf A98}, 592. \\ 
Hamamoto I. (1965). {\it Nucl. Phys.} {\bf 62}, 49.\\
Hannawald M. {\it et al.\/} (2000). {\it Phys. Rev.} {\bf C62}, 054301. \\ 
Hirsch M., Staudt A., Klapdor-Kleingrothaus H.-V. (1992). {\it Atomic and
   Nucl. Data Tables} {\bf 51}, 244. \\ 
Hirsch M., Staudt A., Klapdor-Kleingrothaus H.-V. (1996). 
  {\it Phys. Rev.} {\bf C54}, 2972. \\ 
K\"oster U. (2000). PhD Thesis, TU M\"unchen, unpublished. \\
Korgul A. {\it et al.\/} (2000). {\it Eur. Phys. J.} {\bf A3}, 167. \\
Kratz K.-L. and Herrmann G. (1973).  {\it Z. Physik} 
   {\bf 263}, 435. \\
Kratz K.-L. (1984). {\it Nucl.Phys.} {\bf A417}, 447. \\ 
Kratz K.-L. {\it et al.\/} (2000). 
   {\it AIP Conf. Proc.} {\bf 529} 295. \\
Krumlinde J. and M\"oller P. (1984). {\it Nucl. Phys.} {\bf A417}, 419.\\
Madland D.M. and Nix J.R. (1988). {\it Nucl. Phys.} {\bf A476}, 1. \\ 
Mach H., Fogelberg B. and Urban W. (2000). priv. comm. and to be 
   published. \\
Mann F.M. {\it et al.\/} (1984). {\it Nucl. Sci. Eng.} {\bf 87}, 418.  \\
Mann F.M. (1986). Specialists' Meeting on Delayed Neutron Properties 
   ISBN 070044 0926 7, p.21. Birmingham, September. \\
Mehren T. {\it et al.\/} (1996).  {\it Phys. Rev. Lett.}
   {\bf 77}, 458. \\
M\"oller P. and Randrup J. (1990). {\it Nucl. Phys.} {\bf A514}, 1.\\
M\"oller P. {\it et al.\/} (1992). {\it Nucl. Phys.} {\bf A536}, 61. \\
M\"oller P. {\it et al.\/} (1995). {\it Atomic Data and Nucl. Data Tables} {\bf 59},
   185. \\
M\"oller P., Nix J.R. and Kratz K.-L. (1997). {\it Atomic Data and Nucl.
   Data Tables} {\bf 66}, 131. \\
Mueller W.F. {\it et al.\/} (2000).  {\it Phys. Rev.} {\bf C61}, 054308. \\
Preston M.A. (1962). ``Physics of the Nucleus'', Addison-Wesley, Reading.\\
Pfeiffer B., Kratz K.-L., M\"oller P. (2000). ``Predictions of $T_{1/2}$ and 
   $P_{\rm n}$ Values''; Internal Report, Inst. f\"ur Kernchemie, 
   Univ. Mainz (2000). \\ 
   URL: www.kernchemie.uni-mainz.de/$\sim$pfeiffer/khf/ \\
Rudstam G. (1993). {\it Atomic Data and Nucl. Data Tables} {\bf 53}, 1. \\
deShalit A. and Feshbach H. (1974). ``Theoretical Nuclear Physics, vol.\ I: 
Nuclear Structure'', Wiley, New York\\
Shergur J. (2000).  Proc. Conf. on {\it Nuclear Structure 2000 -
   NS2000}, MSU August 2000; {\it Nucl. Phys. A}, in print. \\
Sorlin O. {\it et al.\/} (1993). {\it Phys.Rev.} {\bf C47},  2941. \\ 
Sorlin O. {\it et al.\/} (2000).  {\it Nucl. Phys.} {\bf A669}, 351. \\
Tachibana T., Nakata H., Yamada M. (1998). {\it AIP Conf. Proc.} 
   {\bf 425}, 495. \\
Takahashi K. (1972). {\it Prog. Theoret. Phys}. {\bf 47}, 1500. \\ 
Takahashi K. {\it et al.\/} (1973). {\it Atomic Data and Nucl. Data Tables} {\bf12},
    101. \\ 
Wang J.C. {\it et al.\/} (1999).  {\it Phys. Lett.} 
   {\bf B454}, 1. \\
Weissman L. {\it et al.\/} (1999).  {\it Phys. Rev.}
   {\bf C59}, 2004. \\
Wilson W.B. and England T.R. (1993).  {\it Prog. Nucl. Energy}, this volume. \\


\end{document}